\documentclass[10pt]{article}
\usepackage{aas2pp4}
\lefthead{Kassim \& Lazio}
\righthead{Low-Frequency VLA Search for Geminga}

\newcommand{\dtd}{\ensuremath{\Delta t_d}}
\newcommand{\dnud}{\ensuremath{\Delta\nu_d}}

\newcommand{\diss}{\ensuremath{\mathrm{DISS}}}

\newcommand{\mjybm}{\mbox{mJy~beam${}^{-1}$}}

\received{1999 September~13}
\revised{1999 October~14}
\accepted{1999 October~20}
\journalid{ApJ}{}
\paperid{E99-5672, kassim.0913}
\ccc{PD~1999}

\begin{document}
\title{Upper Limits on the Continuum Emission from Geminga at~74
	and~326~MHz}
\author{Namir~E.~Kassim \& T.~Joseph~W.~Lazio}
\affil{Code~7213, Naval Research Laboratory, Washington, DC,
20375-5351; kassim@rsd.nrl.navy.mil; lazio@rsd.nrl.navy.mil}

\begin{abstract}
We report a search for radio continuum emission from the gamma-ray
pulsar Geminga.  We have used the VLA to image the location of the
optical counterpart of Geminga at~74 and 326~MHz.  We detect no radio
counterpart.  We derive upper limits to the pulse-averaged flux
density of Geminga, taking diffractive scintillation into account.
We find that diffractive scintillation is probably quenched at~74~MHz
and does not influence our upper limit, $S< 56$~mJy ($2\sigma$), but
that a 95\% confidence level at~326~MHz is $S < 5$~mJy.  Owing to
uncertainties on the other low-frequency detections and the
possibility of intrinsic variability or extrinsic variability
(refractive interstellar scintillation) or both, our non-detections
are nominally consistent with these previous detections.
\end{abstract}


\section{Introduction}\label{sec:intro}

The gamma-ray pulsar Geminga (PSR~J0633$+$1746, 1E~0630$+$178,
1CG~195$+$04) was first identified as a compact gamma-ray source in
the Galactic anticenter (\cite{hermsenetal77}; \cite{tfhkl77}).
Though initial speculation included the possibility that this source
was a pulsar, its identification as such was not secure until the
detection of X-ray pulsations (\cite{hh92}; \cite{bertschetal92}).
Since pulsars are typically bright in the radio portion of the
spectrum, intense effort has been devoted to finding a radio
counterpart to Geminga (Table~\ref{tab:log}).

\begin{deluxetable}{lcc}
%
% Table of previous limits on Geminga
%
\tablewidth{490pt}
\tablecaption{Observational Limits on the Radio Counterpart of
	Geminga Below~1000~MHz\label{tab:log}}
\tablehead{\colhead{Frequency} & \colhead{Flux Density} &
	\colhead{Reference} \\
	\colhead{(MHz)} & \colhead{(mJy)}}

\startdata
%\cutinhead{Periodicity Searches} % why doesn't \cutinhead work?
\multicolumn{3}{c}{Periodicity Searches} \nl

\phn\phn35\phd\phn & $<$ 100 & 9 \nl % Ramachandran et al.
\phn\phn41\phd\phn & 300     & 7 \nl % Shitov et al.
\phn\phn61\phd\phn & 111     & 7 \nl % Shitov et al.
\phn102\phd\phn    & $<$ 100 & 5 \nl % Kuz'min & Losovskii
\phn102.5          & $60 \pm 95$ & 8 \nl % Malofeev & Malov
                   & (5--500) & \nl

\nl

\phn102.5          & $8^{+3}_{-2}$ & 6 \nl % Shitov & Pugachev
\phn103\phd\phn    & 1000    & 12 \nl % Vats et al.
\phn318\phd\phn    & $<$ 1   & 3 \nl % D'Amico
\phn318\phd\phn    & $<$ 3.1 & 4 \nl % Fauci et al.
\phn318\phd\phn    & $<$ 0.1 & 11 \nl % Buderi et al.

\nl

\phn325\phd\phn    & $<$ 2500 & 1 \nl % Mandolesi et al.
\phn327\phd\phn    & $<$ 0.3 & 9 \nl % Ramachandran et al.
\phn327\phd\phn    & $<$ 3   & 10 \nl % McLaughlin et al.
\phn430\phd\phn    & $<$ 1   & 3 \nl % D'Amico
\phn430\phd\phn    & $<$ 0.6 & 4 \nl % Fauci et al.

\nl 

\phn430\phd\phn    & $<$ 0.05 & 11 \nl % Buderi et al.
\phn928\phd\phn    & $<$ 6   & 2 \nl % Seiradadis (1981)

%\cutinhead{Continuum Searches} % why doesn't \cutinhead work?
\multicolumn{3}{c}{Continuum Searches} \nl

\phn\phn74\phd\phn & $<$ 56 & 13 \nl
\phn326\phd\phn    & $<$ 5.0  & 13 \nl
\enddata

\tablerefs{
(1)~Mandolesi et al.~(1978); 
(2)~Seiradakis~(1981);
(3)~D'Amico~(1983); 
(4)~Fauci et al.~(1984); 
(5)~Kuz'min \& Losovskii~(1997b); 
(6)~Shitov \& Pugachev~(1997); 
(7)~Shitov et al.~(1997); 
(8)~Malofeev \& Malov~(1997); 
(9)~Ramachandran et al.~(1998); 
(10)~McLaughlin et al.~(1999); 
(11)~Buderi et al.~(1999);
(12)~Vats et al.~(1999);
(13)~this work}
\tablecomments{Only the work of McLaughlin et al.~(1999) and that reported 
here have taken interstellar scintillation into account explicitly in 
setting upper limits.}
\end{deluxetable}

Until recently, only upper limits on the radio emission from Geminga
could be established.  However, a number of groups have now reported
detecting pulsed radio emission from Geminga near~100~MHz
(\cite{kl97a}; \cite{kl97b}; \cite{mm97};
\cite{smmp97}; \cite{sp97}; \cite{vdssiod97}; \cite{vsdiosd99}).
However, subsequent attempts to detect pulsed emission at frequencies
of~35 and~326~MHz have failed (Ramachandran, Deshpande, \&
Indrani~1998; \cite{mchm99}).  In an effort to constrain further the
radio spectrum of Geminga, we have conducted 74 and 326~MHz
observations of it.  In \S\ref{sec:observe} we describe our
observations, in \S\ref{sec:scintillate} we obtain upper limits for
the Geminga pulsar that take interstellar scintillation into account,
and in \S\ref{sec:conclude} we discuss our results.

\section{Observations}\label{sec:observe}

Reported previous detections of Geminga have been in the range
41--103~MHz (Table~\ref{tab:log}).  Subsequent attempts to confirm
these detections have been close to, but outside of, this frequency
range.  The non-detection at~35~MHz (\cite{rdi98}) seems problematic,
but a combination of a low-frequency cutoff in the spectrum and
interstellar scintillation may explain the detection at~41~MHz and the
upper limit at~35~MHz.  The observations reported here are significant
in that they are the only attempt to confirm the radio emission from
Geminga within the frequency range of the reported
detections.

Our observations were conducted on 1998 March~9 with the VLA in the A
configuration; Table~\ref{tab:observelog} summarizes various details.
Ionospheric phase variations would have vitiated a periodicity search,
so our observations were designed only to image the pulse-averaged
emission from the pulsar.  The various detections suggest that the
radio pulse is extremely wide, covering perhaps 90\arcdeg\ of pulse
phase.  Thus, our attempt to detect Geminga should not be affected
significantly by searching for a point source at its location.

\begin{deluxetable}{ccccccc}
\tablecaption{VLA Observing Log\label{tab:observelog}}
\tablehead{
 \colhead{$\nu$} & \colhead{$B$} & \colhead{$B_I$} & \colhead{$T$} &
 \colhead{Beam} & \colhead{$\Delta I$} & 
 \colhead{$F$} \nl
 \colhead{(MHz)} & \colhead{(MHz)} & \colhead{(kHz)} & \colhead{(hr)} &
 \colhead{(\arcsec\ $\times$ \arcsec)} & \colhead{(mJy~beam${}^{-1}$)} &
 \colhead{(mJy)}
}
\startdata

\phn74 & 1.5       & 122 & 5.4 & $25 \times 23$ @ 64\arcdeg\   & 28\phd\phn & $-13$\phd\phn\phn \nl
326    & 3\phd\phn & 684 & 5.4 & $5.1 \times 4.8$ @ 75\arcdeg\ & 1.4        & \phs\phn0.28 \nl

\enddata

\tablecomments{$B$ is the total observing bandwidth while $B_I$ is the 
bandwidth used in imaging.  $T$ is the on-source integration time.
$\Delta I$ is the rms noise level in the image, and $F$ is the
measured flux density at the location of Geminga.}
\end{deluxetable}

Post-processing of low-frequency VLA data uses procedures similar to
those at higher frequencies, though certain details differ.\footnote{
A full description of low-frequency VLA data reduction procedures is at
$\langle$URL:\hfill\linebreak
http://rsd-www.nrl.navy.mil/7213/lazio/tutorial/$\rangle$.}
At~74~MHz Cygnus~A served as both the flux density and visibility
phase calibrator.  At~326~MHz the primary flux density calibrator was
3C~48 whose flux density was taken to be 52.2~Jy, and phase
calibration was accomplished with frequent observations of 4C~14.18.
At both frequencies, after initial phase calibration, several
iterations of hybrid mapping were used to improve the dynamic range.

Two significant differences for the post-processing were the impact of
radio frequency interference (RFI) and the large fields of view.  In
order to combat RFI, the data were acquired with a much higher
spectral resolution than used for imaging.  Excision of potential RFI
is performed on a per-baseline basis for each 10-s visibility
spectrum.  The large fields of view (11\arcdeg\ at~74~MHz, 2\fdg5
at~326~MHz) mean that the sky cannot be approximated as flat.  In
order to approach the thermal-noise limit, we used a polyhedron
algorithm (\cite{cp92}) in which the sky is approximated by many
two-dimensional ``facets.''

We did not image the data at the same spectral resolution with which
they were acquired.  Sources in our images are identified on the basis
of their brightness.  Bandwidth smearing can cause weak sources in the
outer portions of the primary beam to have a brightness below our
detection threshold.  The sidelobes of any such undetected sources
contribute to the noise in the image.  Increasing the spectral
resolution has the beneficial result of decreasing the amount of
bandwidth smearing, however, it also increases the computational
expense and the impact of diffractive scintillation. In fact, for
reasonable choices of the scintillation properties for the line of
sight toward Geminga (\S\ref{sec:scintillate}), the impact of
diffractive scintillation will be unchanged for the range
of spectral resolutions we have available.  The primary factors for
deciding the spectral resolution are the desire to minimize bandwidth
smearing \textit{vs.}  computational expense.  The bandwidths we
used---122~kHz at~74~MHz and~684~kHz at~326~MHz---gave modest spectral
resolutions (a few), modest bandwidth smearing near the edge of the
primary beam ($\sim 10$\%), and reasonable imaging times (less than a
few days of computation).

Figure~\ref{fig:74} shows our 74~MHz image of the immediate region
around Geminga.  The off-source rms noise level on this image is
28~\mjybm, and the estimated thermal limit is 25~\mjybm.
Figure~\ref{fig:326} shows our 326~MHz image of the immediate region
around Geminga.  The off-source rms noise level on this image is
1.4~\mjybm, and the estimated thermal limit is 0.5~\mjybm.

One additional impact on the 74~MHz observations is ionospheric
refraction, which shifts the field of view without distorting the
brightness distribution (\cite{e84}). Shifts at~74~MHz are typically a
few arcminutes and vary on time scales of tens of minutes.
Self-calibration ``freezes'' out this refraction but destroys absolute
position information.  Fortunately, the primary beam is large enough
that a typical image contains many tens of sources identified in the
1400~MHz NRAO VLA Sky Survey (\cite{ccgyptb98}).  We constructed a
reference grid using 42 NVSS sources that were either unresolved or
slightly resolved (diameters $< 30\arcsec$).  In determining the
reference grid, we checked for any systematic trends as a function of
position within the 74~MHz image, flux densities of the sources, or
angular diameters of the sources.  None were found.  We shifted the
74~MHz image by the median offsets, 3\farcs9 in right ascension and
$-58\farcs4$ in declination.  Based on the variances of the offsets
from the reference sources, our 74~MHz astrometry should be accurate
to approximately 5\arcsec, a fraction of the beam diameter, for an
individual source.  Figure~\ref{fig:74} has been corrected for
refraction.

\section{Upper Limits and Interstellar Scintillation}\label{sec:scintillate}

In this section we address our upper limits on Geminga's flux density
and the impact of scintillation on them.  We shall follow the
procedure described by McLaughlin et al.~(1999) for dealing with
the effects of interstellar scintillation (ISS).  

Briefly, their procedure involves convolving the known probability
density functions (pdfs) for the diffractive ISS gain and
noise in the image to produce a pdf for the measured flux density for
the scintillating source.  This pdf is then multiplied by a
\textit{prior} for the intrinsic flux density of the source (Geminga)
to form the \textit{posterior} for the intrinsic flux density of the
source.  As our images were constructed using an FFT, without
specifying a zero-spacing flux density, we have assumed (and verified)
the noise in the image has a zero-mean Gaussian pdf.

In addition to the rms noise in the image
(Table~\ref{tab:observelog}), we require two additional parameters: a
measure of the strength of diffractive ISS and a measured flux density
at the location of Geminga.  The former is provided by $N_\diss$, the
number of ``scintilles'' or scintillation intensity maxima occurring
during the entire observation (\cite{c86};
\cite{mchm99}).  For Geminga, there are no published measurements of
the relevant ISS parameters, the scintillation time~\dtd\ or the
scintillation bandwidth~\dnud.  McLaughlin et al.~(1999) estimated
these parameters to be $\dtd \approx
275\,\mathrm{s}\,(\nu/327\,\mathrm{MHz})^{2.2}$ and $\dnud \approx
1.5\,\mathrm{MHz}\,(\nu/327\,\mathrm{MHz})^{4.4}$.  Their values imply
a scattering measure toward Geminga of $\mathrm{SM} \sim
10^{-4.4}$~kpc~m${}^{-20/3}$.  While consistent with the local
distribution of scattering material, there are also large variations
in the value of SM for pulsars at distances comparable to that of
Geminga: PSR~J0437$-$4715 with $D = 0.18$~kpc and $\mathrm{SM} =
10^{-3.8}$~kpc~m${}^{-20/3}$ (\cite{sbmnka97}; Johnston, Nicastro,
\& Koribalski~1998) \textit{vs.}  PSR~B0950$+$08 with $D = 0.125$~kpc
and $\mathrm{SM} = 10^{-4.5}$~kpc~m${}^{-20/3}$ (\cite{gtwr86};
\cite{pc92}).  We shall adopt McLaughlin et al.'s~(1999) value for SM
but note how a different value affects our results.  For our
integration time and receiver bandwidth, the resulting values are
$N_\diss = 910$ at~74~MHz and $N_\diss = 15$ at~326~MHz.

We obtain the flux density at the location of Geminga by integrating
over an area of~1~beam at the position of Geminga.  We use
the term ``flux density at the location of Geminga'' because of the
noise in the image.  As the noise is zero-mean, there are both
positive and negative intensity variations.  A weak source combined
with a large negative noise deviation could result in a measured flux
density quite different than that of the intrinsic flux density of the
source.  Given the sizes of the beams (6\arcsec\ at~326~MHz
and~25\arcsec\ at~74~MHz), this integration also takes into account
the uncertainty in the position of Geminga which is less than
1\arcsec\ (\cite{bcm93}; \cite{cbmt96}).

\subsection{74~MHz}

We begin by determining the upper limit on Geminga's flux density if
ISS was not an issue.  The position of Geminga's optical counterpart,
even taking into account its proper motion, is known to accuracy far
better than the beam diameter of~25\arcsec.  We therefore adopt an
upper limit of~56~mJy, twice the rms noise level.

As $N_\diss \to \infty$, DISS becomes more and more heavily quenched,
and the pdf of the DISS gain approaches a $\delta$-function centered
on unity.  We consider $N_\diss \sim 10^3$ sufficiently large that
DISS will not affect our upper limit.  If our adopted value of SM were
too large, and the correct value were $\mathrm{SM} \sim
10^{-5.3}$~kpc~m${}^{-20/3}$---comparable to that for
PSR~B0950$+$08---then $N_\diss = 44$.  We would obtain an upper limit
for Geminga's flux density of $S < 86$~mJy at the 95\% confidence
level.

\subsection{326~MHz}

If we ignore ISS, the 6\arcsec\ beam remains large enough that we
adopt an upper limit of~2.8~mJy, again twice the rms noise level.

Our adopted value of SM implies $N_\diss = 15$.  In contrast to the
situation at~74~MHz, DISS can have a substantial impact on the adopted
flux density at~326~MHz.  We find an upper limit on Geminga's flux
density of $S < 5.0$~mJy at the 95\% confidence level.  If our adopted
value of SM were too large, and the correct value were $\mathrm{SM}
\sim 10^{-5.3}$~kpc~m${}^{-20/3}$, then $N_\diss = 5$.  We would
obtain an upper limit for Geminga's flux density of $S < 13.5$~mJy at
the 95\% confidence level.

\section{Conclusions}\label{sec:conclude}

Figure~\ref{fig:spectrum} incorporates our upper limits with the
limits and detections of Geminga below~1000~MHz in the literature;
Table~\ref{tab:log} summarizes the data shown in
Figure~\ref{fig:spectrum}.

Can we reconcile our nondetection with the detections near~100~MHz?
One simple possibility is that the pulse-averaged radio emission from
Geminga is intrinsically time variable.  In this case, our
nondetection is quite easy to reconcile with the various detections.
Vats et al.~(1999) argue that Geminga is intrinsically time variable
in order to explain their detection with a flux density of~1~Jy
at~103~MHz, and they speculate on intrinsic mechanisms that could be
responsible.

Even if we treat the Vats et al.~(1999) detection at~103~MHz as an
outlier, though, our nondetection is just consistent with the other
detections near~100~MHz.  Shitov \& Pugachev~(1997) report a flux
density of $8^{+3}_{-2}$~mJy at~102.5~MHz.  Clearly, if this is the
intrinsic flux density, then there is no difficulty reconciling this
with our nondetection.

There are two difficulties with assuming a flux density of~8~mJy
at~102~MHz, though.  First, it requires Geminga to have an extremely
steep spectrum between 61 and~102~MHz, $\alpha \approx -5$, though,
Geminga's radio emission mechanism might be different than that of
other pulsars.  Second, it is difficult to explain other detections
at~102~MHz.  Some combination of intrinsic and ISS-induced intensity
variation could in principle explain this dynamic range, but, in
practice it might require improbably large DISS gains.

A more likely value for the 100~MHz flux density is near 50~mJy, in
which case Shitov \& Pugachev~(1997) detected it in an ISS-depressed
state.  This explanation is appealing in two respects.  First, a DISS
gain of~0.1 would depress the flux density from near 50~mJy to near
5~mJy, and a DISS gain this small has a probability of occurrence near
90\%.  Second, the spectral index between 61 and~102~MHz would be
$-1.6$, more characteristic of radio pulsars.  In this case, the
expected flux density at~74~MHz is 80~mJy, above our upper limit for
our nominal value of $N_\diss$, but below it if our adopted value of
SM is too large.

In reconciling our nondetection at~74~MHz with the detections, we have
relied on the robustness of the 41 and~61~MHz observations.  In
reporting their detections at these frequencies, Shitov et al.~(1997)
did not indicate to what extent DISS had been quenched.  As DISS is
more likely to depress rather than enhance the flux density of a
source, the 41 and~61~MHz detections could represent \emph{lower}
limits to Geminga's flux density.  If so, reconciliation would be more
difficult.

One aspect of ISS not addressed quantitatively is the impact of
\emph{refractive} \hbox{ISS}.  At low frequencies, ISS has an ancipital
nature: DISS occurring on ``short'' time scales and RISS occurring on
``long'' time scales.  RISS is more difficult to quench than DISS
because its time scales are typically hours to days, so the flux
density of Geminga could be enhanced or suppressed for an entire
observing run.  RISS could serve as the extrinsic mechanism that would
contribute both to our non-detection (and that of \cite{rdi98}
at~35~MHz) and the detections near~100~MHz.

RISS is more difficult to handle quantitatively because its pdf is not
as well constrained as the pdf for \hbox{DISS}.  Some indication of
the effect of RISS can be obtained by adopting a log-normal pdf
(\cite[Chap.~20]{i78}).  Gupta, Rickett, \& Coles~(1993) found a
refractive intensity modulation variance $m_r^2 \approx 0.2$ for
PSR~B0950$+$08 at~74~MHz.  This variance, when used with a log-normal
pdf, implies a (95\% probability) range of RISS-induced variations of
a factor of a few.  Thus, these low-frequency measurements and upper
limits could have an additional factor of two or so uncertainty.

In summary, we have not detected Geminga at 74~MHz ($S < 56$~mJy)
or~326~MHz ($S < 5$~mJy).  We can reconcile our nondetection with
previous low-frequency detections by invoking intrinsic variability or
extrinsic variability (refractive interstellar scintillation) or both.

\acknowledgements
This research made use of NASA's ADS Abstract Service and the SIMBAD
database, operated at the CDS, Strasbourg, France.  The National Radio
Astronomy Observatory is a facility of the National Science Foundation
operated under cooperative agreement by Associated Universities, Inc.
The results presented here made use of the Department of Defense High
Performance Computing Modernization Program.  Basic research in radio
astronomy at the NRL is supported by the Office of Naval Research.

\clearpage
\onecolumn

\begin{figure}
\caption[]{The Geminga field at~74~MHz.  The cross marks the location of
the optical counterpart of Geminga (\cite{clmmmpb98}); the uncertainty 
in the location of Geminga is approximately 100 times smaller than the 
size of the cross.  The beam is $24\farcs6 \times 23\farcs3$ at a
position angle of~64\arcdeg, and the rms noise level is 28~\mjybm.
Contour levels are 28~\mjybm\ $\times$ $-2$, 2, 3, 5.  Also indicated
are two sources from the NRAO VLA Sky Survey (\cite{ccgyptb98}).}
\label{fig:74}
\end{figure}

\begin{figure}
\caption[]{The Geminga field at~326~MHz.  A cross marks the location
of the optical counterpart of Geminga (\cite{clmmmpb98}); the uncertainty 
in the location of Geminga is approximately 100 times smaller than the 
size of the cross.  The beam is $5\farcs \times 4\farcs8$ at a
position angle of~75\arcdeg, and the rms noise level is 1.4~\mjybm.
Contour levels are 1.4~\mjybm\ $\times$ $-3$, 3, 3, 5, 7.07, 10, 14.1,
$\ldots$.  Also indicated are a number of sources from the NRAO VLA
Sky Survey (\cite{ccgyptb98}).}
\label{fig:326}
\end{figure}

\begin{figure}
\caption[]{The radio spectrum of Geminga below 1000~MHz.  The
upper limits reported in this paper are marked with a star.  Limits
and detections are summarized in Table~\ref{tab:log}.  With the
exception of the upper limit by McLaughlin et al.~(1999; $< 3$~mJy
at~326~MHz), all other upper limits have not considered the effects of
interstellar scintillation.}
\label{fig:spectrum}
\end{figure}

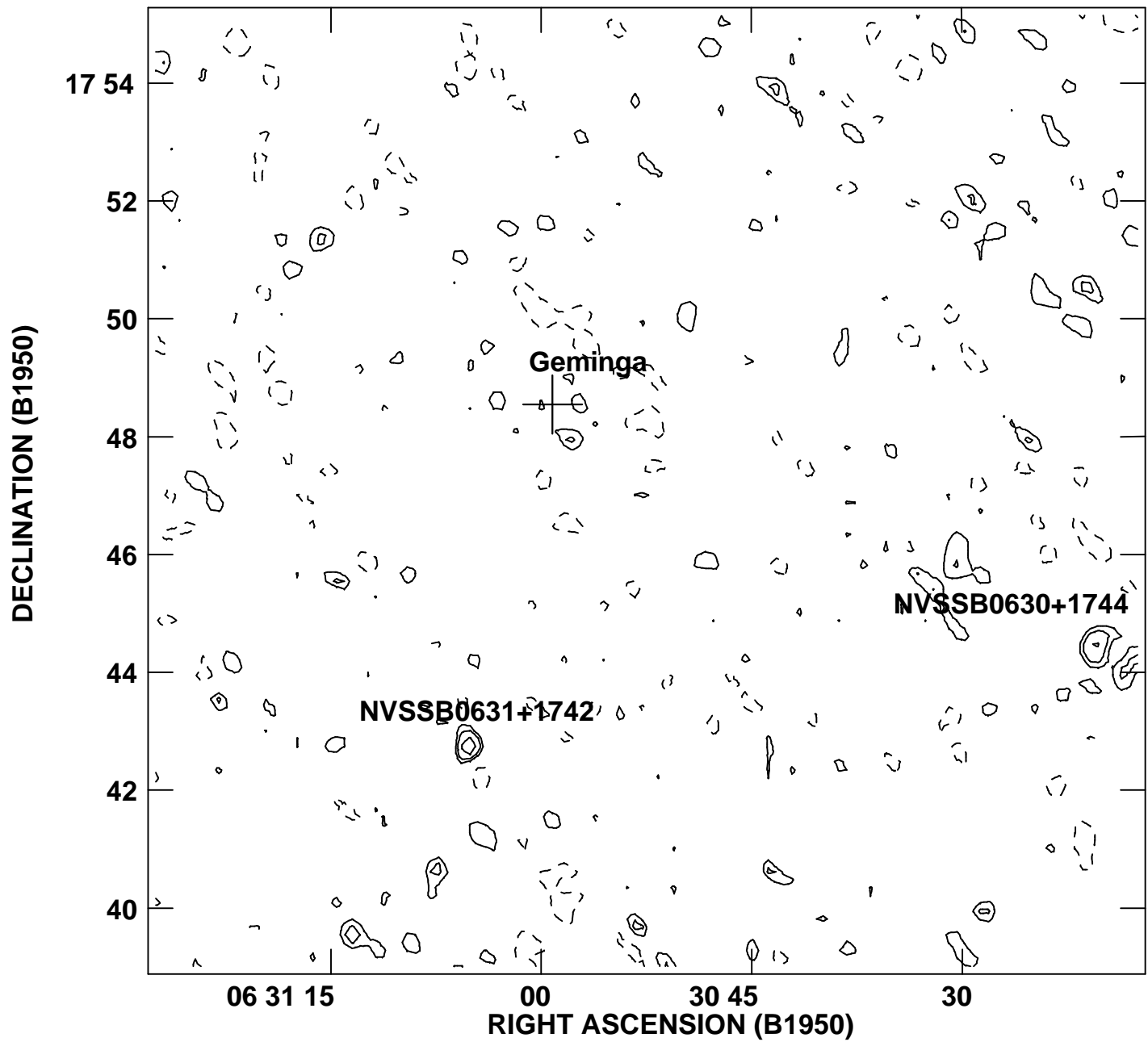
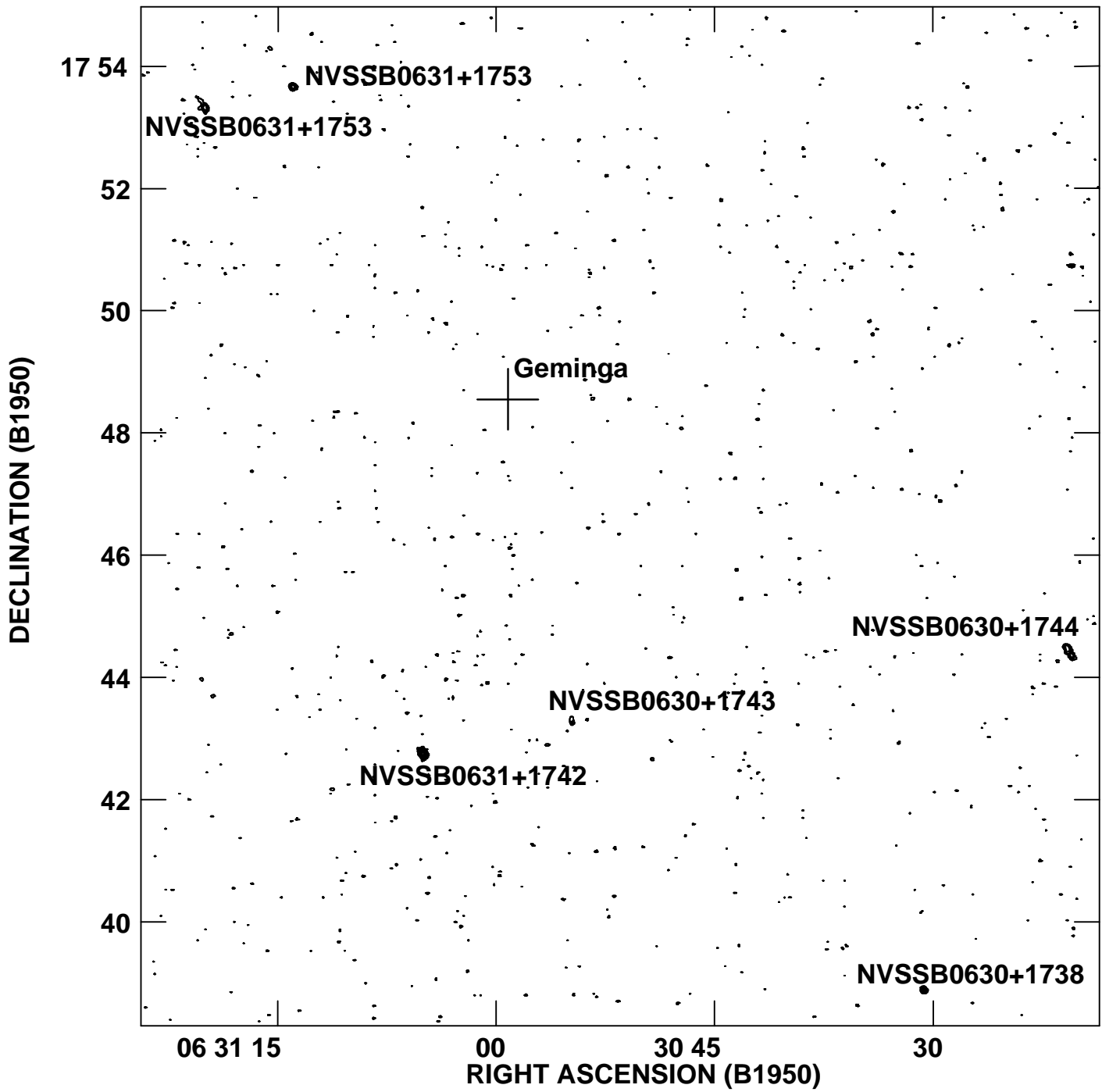
\end{document}